\documentclass{owrart}
\usepackage[latin1]{inputenc}  
\newcommand{\href}[2]{\texttt{#2}}
\usepackage[dvipdfm]{graphicx,color} 
\graphicspath{{dessins/}{./}}

\begin{document}


\begin{talk}[A. Bobenko, Y. Suris]{Christian Mercat}
{Discrete Riemann surfaces, linear and non-linear}
{Mercat, Christian}

To the first two orders, discrete holomorphic functions are whether
complex linear or Möbius transformations. We discretize this notion to
complex functions of vertices $\diamondsuit_0$ of an oriented
quad-mesh $\diamondsuit$. First, fix a direct embedding of this
quad-mesh into the complex plane by a complex function of the vertices
$Z:\diamondsuit_0\to C$, corresponding to the identity map. We will
say that another function is \emph{linear holomorphic} (with respect
to $Z$) if and only if the ratio along the diagonals are the same as
$Z$ on each quadrilateral:
$$\forall (x,y,x',y')\in\diamondsuit_2,\;
\frac{f(y')-f(y)}{f(x')-f(x)}=\frac{Z(y')-Z(y)}{Z(x')-Z(x)}=i\,\rho_{(x,x')}.$$
Likewise we will say that a function is \emph{cross-ratio preserving}
if the cross-ratio on each quadrilateral is the same as the one given
by $Z$:
$$\frac{f(x)-f(y)}{f(y)-f(x')}\frac{f(x')-f(y')}{f(y')-f(x)}=\frac{Z(x)-Z(y)}{Z(y)-Z(x')}\frac{Z(x')-Z(y')}{Z(y')-Z(x)}=q_{(x,y,x',y')}.$$
An important class of cross-ratio preserving maps are given by circle
patterns with prescribed intersection angles. In the form of a
\emph{Hirota system}, a connection between the two notions can be
described: A discrete function $F$ is cross-ratio preserving if its
exterior differential can be written, on each edge
$(x,y)\in\diamondsuit_1$ as $F(y)-F(x)=f(x)f(y)\bigl(Z(y)-Z(x)\bigr)
=:\int_{(x,y)}f\,dZ$ for a function $f$. The constraint on $f$ is the
Morera equation: $\oint f\,dZ=0$ around every quadrilateral. Looking
at logarithmic derivatives $f_\epsilon=f\times(1+\epsilon g)$ of $f$
that still form a Hirota system, one finds that $g$ should be linear
holomorphic with respect to $F$: $\frac{g(y')-g(y)}{g(x')-g(x)}=\frac{F(y')-F(y)}{F(x')-F(x)}.$
Linear constraints can as well be reformulated as a Morera equation
$\oint f\, dZ=0$ not for a multiplicative but for an additive coupling
between functions and $1$-forms: 
$\int_{(x,y)}f\, dZ:=\frac{f(x)+f(y)}2\bigl(Z(y)-Z(x)\bigr).$

We will be interested in so called \emph{critical} reference maps $Z$
composed of rhombi, the length $\delta=|Z(y)-Z(x)|$ is constant for
all edges $(x,y)\in\diamondsuit_1$. Then these two constraints are
\emph{integrable}, meaning that they give rise to a well-defined
Bäcklund (or Darboux) transform: given a linear holomorphic resp.
cross-ratio preserving map $f$ and a starting point
$O\in\diamondsuit_0$, one can define a $2$-parameters family
$f_{\lambda,\mu}$ of deformations of $f$ that still fulfill the same
(linear resp.  cross-ratio) condition and starting value $\mu$ at the
point $O$. This is done by viewing the solution $f$ lying on a ground
level and building ``vertically'', above each edge $(x,y)$, a
quadrilateral on which the same kind of equation will be imposed:
$\frac{f_{\lambda,\mu}(x)-f(y)}{f_{\lambda,\mu}(y)-f(x)}=\frac{\lambda+Z(x)-Z(y)}{\lambda+Z(y)-Z(x)}$
for the linear constraint, resp.
$\frac{f_{\lambda,\mu}(x)-f(x)}{f(x)-f(y)}
\frac{f(y)-f_{\lambda,\mu}(y)}{f_{\lambda,\mu}(y)-f_{\lambda,\mu}(x)}
=\frac{\lambda^2}{\bigl(Z(y)-Z(x)\bigr)^2}$ for the cross-ratio
constraint.

In~\cite{CM:BMS}, based on the point of view of integrable systems, we
define discrete holomorphicity in $\mathbb{Z}^d$, for $d>1$ finite,
equipped with rapidities $(\alpha_i)_{1\leq i\leq d}$ and the tools of
integrable theory yields interesting results like a \emph{Lax pair}
governing a {\em moving frame} $\Psi(\cdot,\lambda):\mathbb{Z}^d\to
GL_2(\mathbb{C})[\lambda]$ and \emph{ isomonodromic} solutions like
the \emph{Green function} found by Kenyon~\cite{CM:Ken02}.

An important tool of the linear theory is the existence of an explicit
basis of discrete holomorphic functions. In the rhombic case, for a
discrete holomorphic function $f$, the $1$-form $f\, dZ$ is
holomorphic (it is closed and its ratios on two dual diagonals are
equal to the reference ratios) and can be integrated, yielding back a
holomorphic function.  Differential equations can be setup, producing
explicit formulae for \emph{exponentials and polynomials} which are
shown to form a basis of discrete holomorphic functions.

Lots of results of the continuous theory can be extended to this
discrete settting: There exists a \emph{Hodge star}: $*:C^k\to
C^{2-k}$, defined by $\int_{(y,y')}*\,\alpha :=
\rho_{(x,x')}\,\int_{(x,x')}\alpha$; the discrete \emph{Laplacian} is
written as usual $\Delta := d\, d^* + d^*d$ with $d^*:=-*d\, *$, which
reads as weighted differences around neighbours $\Delta\, f(x) = \sum
\rho_{(x,x_k)}(f(x)-f(x_k)).$ The weights are given by the usual cotan
formula.

A wedge product defines an $L^2$ norm for functions and forms by
$(\alpha,\beta):=\iint_{\diamondsuit_2} \alpha\wedge *\, \bar\beta$.
The norm of $df$ is called the \emph{Dirichlet} energy of the function
$f$, $E_D(f):=\lVert df\rVert^2=
\left(df,\,df\right)=\frac12\sum_{(x,x')\in\Lambda_1}\rho(x,x')
\left\lvert f(x') - f(x) \right\rvert^2.$ The \emph{conformal energy}
of a map measures its conformality defect $E_C(f) := \tfrac12\lVert df
-i * df\rVert^2.$ They are related through $E_C(f) =E_D(f) - 2
\mathcal{A}(f)$ just as in the continuous.

The Hodge star decomposes forms into exact, coexact and harmonic ones,
the harmonic being the orthogonal sum of holomorphic and
anti-holormorphic ones. A Weyl's lemma and a Green's identity are
found.

Non closed $1$-forms with prescribed diagonal ratios define
\emph{meromorphic} forms and the holonomy around a quadrilateral is
called its \emph{residue}.  The compact case is covered with flat
atlases of critical maps for a given euclidean metric with conic
singularities.  \emph{Meromorphic} forms of prescribed holonomies and
poles are defined and are used to form a basis of the space of
holomorphic forms. It is $2g$-dimensional on a genus $g$ surface, that
is \emph{twice} as large as the continuous case, defining two period
matrices instead of one. This difference is explained by the doubling
of degrees of freedom, and partially solved through continuous limit
theorems: the two period matrices converge to the same limit when
refinements of quad-meshes for a given Euclidean metric with conic
singularities are taken. Every holomorphic function can be
approximated by a converging sequence of discrete holomorphic
functions on refinements of critical quad-meshes.

The Green function and potential allow one to setup a Cauchy integral
formula giving the value at a point (in fact its average at two
neighbours $x,y$) as a contour integral: $\oint_{\partial D} f \,
dG_{x,y} = 2 i \pi \,\frac{f(x) + f(y)}{2}.$

We define a derivation with respect to $Z$ by
$\partial:C^0(\diamondsuit)\to C^2(\diamondsuit)$ with $\partial f =
\bigl[(x,y,x',y')\mapsto -\frac
i{2\mathcal{A}(x,y,x',y')}{\displaystyle\oint\limits_{(x,y,x',y')}f
  d\bar Z}\bigr]$ $=\frac{(f(x')-f(x))(\bar y'-\bar y)-(\bar x'-\bar
  x)(f(y')-f(y))}{(x'-x)(\bar y'-\bar y)-(\bar x'-\bar x)(y'-y)},$
where $Z(x)$ is simply written $x$; and likewise $\bar\partial f$ with
$f d Z$. A holomorphic function $f$ verifies $\bar\partial f\equiv 0$
and $\partial f(x,y,x',y') = \frac{f(y')-f(y)}{y'-y} =
\frac{f(x')-f(x)}{x'-x}.$ The Jacobian \hbox{$J={|\partial
    f|^2-|\bar\partial f|^2}$} relates the areas
$\smash{\iint\limits_{(x,y,x',y')}} df\wedge\overline{df}= J
\smash{\iint\limits_{(x,y,x',y')}} dZ\wedge\overline{dZ}.$\\[.2em]

\begin{minipage}{0.7\linewidth}
  Following Colin de Verdière and Kenyon, a geometrical interpretation
  of linear discrete holomorphicity is enlightening. As circle
  patterns with prescribed angles can be checked by eye, so can be a
  linear holomorphic map: The quad-mesh $\diamondsuit$, when
  bipartite, decomposes into two dual graphs $\Gamma$ and $\Gamma^*$
  whose edges are dual diagonals of each quadrilateral.  Around each
  vertex $x\in\Gamma_0$, there is a polygon, image of the dual face
  $x^*\in\Gamma^*_2$ by the reference map $Z$. Consider the identity
  map as a picture of all these polygons shrunk by a factor half. It
  represents both dual graphs at the same time as matching polygons. A
  map $f:\diamondsuit_0\to\mathbb{C}$ is \emph{discrete holomorphic}
  if and only if every polygon $x^*$, centered at $f(x)$, scaled and
  turned according to $\partial f (x)$, form into a polygonal pattern
  of the same combinatorics as the reference polygonal pattern, made
  of \emph{similar} polygons.
\end{minipage}
\begin{minipage}{0.3\linewidth}
    \centering
    \includegraphics[width=5cm]{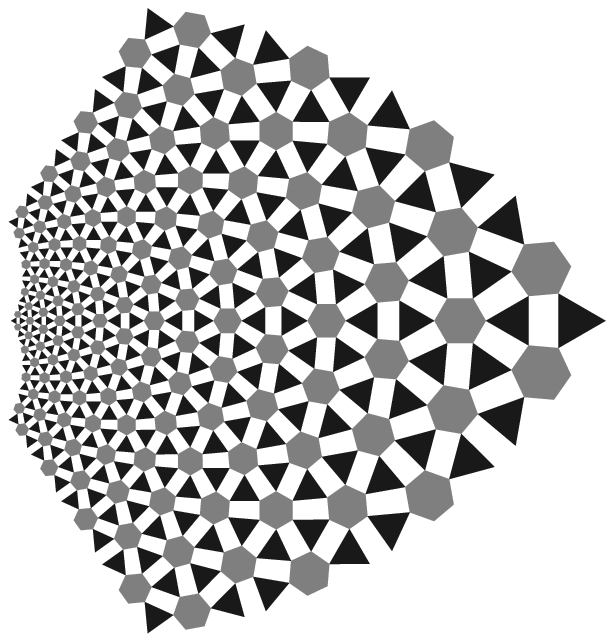}\\
{\small The discrete exponential as a polygonal
      pattern on the triangular/hexagonal lattice}
\end{minipage}
The \emph{dilatation} coefficient of a discrete map $f$ is defined as
$D_f := \frac{|\partial f|+|\bar\partial f|}{|\partial f|-|\bar\partial f|}$. We will call
$f$ \emph{quasi-conformal} when $D_f\geq 1$, that is ${|f_{\bar
    z}|\leq |\partial f|}$. It can be written in term of the \emph{complex
  dilatation}: $\mu_f = \frac{\bar\partial f}{\partial f} = \frac{(f(x')-f(x))(
  y'- y)-( x'- x)(f(y')-f(y))}{(f(x')-f(x))(\bar y'-\bar y)-(\bar
  x'-\bar x)(f(y')-f(y))}.$

\end{talk}


\begin{thebibliography}{10}

\bibitem{CM:M}
Christian Mercat.
\newblock {\em Holomorphie discr\`ete et mod\`ele d'Ising}.
\newblock PhD thesis, Universit\'e Louis Pasteur, Strasbourg, France, 1998.
\newblock under the direction of Daniel Bennequin, Pr\'epublication de l'IRMA,
  available at
  \href{http://tel.archives-ouvertes.fr/tel-00001851/}%
  {http://tel.archives-ouvertes.fr/tel-00001851/}.

\bibitem{CM:M01}
Christian Mercat.
\newblock Discrete {R}iemann surfaces and the {I}sing model.
\newblock {\em Comm. Math. Phys.}, 218(1):177--216, 2001.

\bibitem{CM:M04}
Christian Mercat.
\newblock Exponentials form a basis of discrete holomorphic functions on a
  compact.
\newblock {\em Bull. Soc. Math. France}, 132(2):305--326, 2004.

\bibitem{CM:Ken02}
R.~Kenyon.
\newblock The {L}aplacian and {D}irac operators on critical planar graphs.
\newblock {\em Invent. Math.}, 150(2):409--439, 2002.
\newblock \href{http://fr.arXiv.org/abs/math-ph/0202018}{\tt math-ph/0202018}.

\bibitem{CM:BMS}
Alexander~I. Bobenko, Christian Mercat, and Yuri~B. Suris.
\newblock Linear and nonlinear theories of discrete analytic functions.
  {I}ntegrable structure and isomonodromic {G}reen's function.
\newblock {\em J. Reine Angew. Math.}, 583:117--161, 2005.

\bibitem{CM:CSMcC}
Ruben Costa-Santos and Barry~M. McCoy.
\newblock Dimers and the critical {I}sing model on lattices of genus {$>1$}.
\newblock {\em Nuclear Phys. B}, 623(3):439--473, 2002.
\newblock \href{http://arXiv.org/abs/hep-th/0109167 }{\tt hep-th/0109167}.

\bibitem{CM:Duf}
R.~J. Duffin.
\newblock Basic properties of discrete analytic functions.
\newblock {\em Duke Math. J.}, 23:335--363, 1956.

\bibitem{CM:Duf68}
R.~J. Duffin.
\newblock Potential theory on a rhombic lattice.
\newblock {\em J. Combinatorial Theory}, 5:258--272, 1968.

\bibitem{CM:LF}
Jacqueline Lelong-Ferrand.
\newblock {\em Repr\'esentation conforme et transformations \`a int\'egrale de
  {D}irichlet born\'ee}.
\newblock Gauthier-Villars, Paris, 1955.

\bibitem{CM:M0111043}
Christian Mercat.
\newblock {Discrete Period Matrices and Related Topics}.
\newblock \href{http://arXiv.org/abs/math-ph/0111043}{\tt math-ph/0111043}.

\end{thebibliography}
\end{document}